\documentclass[twoside,leqno,twocolumn]{article}
\usepackage{ltexpprt}
\usepackage[utf8]{inputenc}
\usepackage{amsmath}
\usepackage{amssymb}
\usepackage{graphicx}
\usepackage{xcolor}
\usepackage{url}
\usepackage{subfig}

\usepackage{multicol}
\usepackage{latexsym,amsmath,amssymb,stmaryrd,mathtools}
\usepackage{graphicx,epsfig,tikz, pgfplots}
\usepackage[normalem]{ulem}
\usepackage{float}
\usetikzlibrary{shapes.geometric,positioning,chains,fit,calc}
\tikzset{near start abs/.style={xshift=1cm},
	node/.style={circle,draw},
	nodeone/.style={circle,draw,gray, line width=0.7mm},
	nodered/.style={circle,draw,red, line width=0.7mm}}
	
\newenvironment{customlegend}[1][]{%
    \begingroup
    \csname pgfplots@init@cleared@structures\endcsname
    \pgfplotsset{#1}%
}{%
    \csname pgfplots@createlegend\endcsname
    \endgroup
}%
	
\def\addlegendimage{\csname pgfplots@addlegendimage\endcsname}

\usepackage[useregional]{datetime2}

\newcommand{\Scol}{{\cal S}}

\newcommand{\parent}{\ensuremath{\textrm{parent}}}

\newcommand*{\myblock}[1]{\subparagraph{#1.}}

\title{Hierarchical Relative Lempel-Ziv Compression\thanks{This work was supported in part by the Academy of Finland via grants 339070 and 351150 and the Danish Research Council via grant DFF-8021-002498.}
}

\author{Philip Bille \\ \texttt{phbi@dtu.dk} 
\and Inge Li G{\o}rtz \\ \texttt{inge@dtu.dk} 
\and Simon J. Puglisi \\ \texttt{simon.puglisi@helsinki.fi}
\and Simon R. Tarnow \\ \texttt{}
}

\date{}

\begin{document}

\maketitle

\begin{abstract}
Relative Lempel-Ziv (RLZ) parsing is a dictionary compression method in which a string $S$ is compressed relative to a second string $R$ (called the reference) by parsing $S$ into a sequence of substrings that occur in $R$. RLZ is particularly effective at compressing sets of strings that have a high degree of similarity to the reference string, such as a set of genomes of individuals from the same species. With the now cheap cost of DNA sequencing, such data sets have become extremely abundant and are rapidly growing. In this paper, instead of using a single reference string for the entire collection, we investigate the use of different reference strings for subsets of the collection, with the aim of improving compression.
In particular, we form a rooted tree (or hierarchy) on the strings and then compressed each string using
RLZ with parent as reference, storing only the root of the tree in plain text. To decompress, we traverse the tree in BFS order starting at the root, decompressing children with respect to their parent. We show that this approach leads to a twofold improvement in compression on bacterial genome data sets, with negligible effect on decompression time compared to the standard single reference approach. We show that an effective hierarchy for a given set of strings can be constructed by computing the optimal arborescence
of a completed weighted digraph of the strings, with weights as the number of phrases in the RLZ parsing
of the source and destination vertices. We further show that instead of computing the complete graph, a
sparse graph derived using locality sensitive hashing can significantly reduce the cost of computing a good hierarchy, without adversely effecting compression performance.
\end{abstract}

\section{Introduction}

Given a collection of $m$ strings $\Scol = \{S_1, S_2, \ldots, S_m \}$ of total length $n$, the relative Lempel-Ziv (RLZ) compression scheme parses each string $S_i$, $i>1$, into a sequence of $z$ substrings of the string $S_1$ (we give a formal definition below). If the strings in $S$ are highly similar, the number of substrings---called phrases---in the parsing is small relative to the total length of the collection. In order to achieve compression, phrases (substrings) are replaced by their starting and ending positions in $S_1$.

\medskip

RLZ is a natural choice for compressing sets of full genome sequences of individuals of the same species~\cite{KPZ10,DG11a,DDG13,VNVPM18,NSMG19}, which, because of the high degree of sequence similarity, tend to result in parsings with relatively few phrases. Enormous reductions over the past two decades in the cost of DNA sequencing has led to large and growing databases containing hundreds of thousands of full genome sequences of strains of many known bacteria and viruses. These databases are key, for example, to the field of genomic epidemiology, to screen patient samples (which are sequenced and compared against the genome database) for known pathogenic strains to arrive at diagnosis and suitable treatments (see, e.g.,~\cite{MKASHMCHH21}). 

For a given bacteria, {\em E.coli} say, a genome database may contain genome sequences of thousands of different strains. While all these strains are relatively similar to each other, some share a higher degree of similarity with genomes in a cluster of related strains than they do with other sequences in the database. With this is mind, while RLZ may result in good compression when a single arbitrary sequence is selected as the reference, intuitively it would seem that even more effective compression of the database could be acheived by selecting a different reference for each cluster of strains.

\medskip

\myblock{Our contribution} In this paper we explore the use of more than one reference sequence in the context of RLZ compression. In particular, we show that arranging the sequencing in $\mathcal{S}$ into a hierarchy can lead to significantly better compression than the use of an arbitrarily selected reference, on viral and bacterial genome sets. Finding a hierarchy for a given set of genome sequences leads us into a number of interesting algorithmic problems. In particular:
\begin{enumerate}
    \item To derive an effective hierarchical arrangement for a given set of strings we compute the optimal arborescence of a complete weighted digraph of the strings, with edge weights assigned as the number of phrases in the RLZ parsing of the source and destination vertices. We show that this scheme leads to a factor of 2 improvement to compression on bacterial genomes, and up to a factor 10 on viral genomes, without adversely affecting the speed of decompression.
    \item While the optimal arboresence leads to pleasing compression improvements, it adds significantly to compression time. We show that by sparsifying the graph via locally sensitive hashing compression time can be kept reasonable, while not sacrificing compression gains. 
    \item Along the way, we describe an efficient implementation of the optimal arborescence algorithm of Tarjan~\cite{Tarjan1977} that uses a two-level heap,
     which may be of independent interest.
\end{enumerate}

In the remainder of this section we review related work on RLZ compression. Section~\ref{sec:basics} then lays out notation and basic concepts used throughout. In Section~\ref{sec:hrlz} we formally define hierarchical relative Lempel-Ziv compression, and then go onto describe efficient methods for computing it in Section~\ref{sec:optimaltree} and Section~\ref{sec:lsh}. Section~\ref{sec:fastarbor} describes our engineering of the arborescence algorithm of Tarjan~\cite{Tarjan1977}. Our experimental results on three genomic data sets are presented in Section~\ref{sec:experiments}, before conclusions and reflections are offered.

\medskip

\myblock{Related Work} To the best of our knowledge, the idea of using a hierarchy of reference sequences in RLZ is new. A related notion is mentioned cryptically in Storer and Szymanski~\cite{SS82}, but appears never to have  been implemented. The closest work we could find in the literature is due to Deorowicz and Grabowski~\cite{DG11a}, who describe an RLZ-based scheme for genome compression in which a sequence is compressed relative to multiple reference sequences, with each phrase storing which reference sequence it is from (see also~\cite{DDG13}). Another more recent hierarchical compression scenario is the persistent strings model~\cite{BG2020}. Both of these are quite different to the hierarchical arrangement of sequences we describe here.  

Beyond genomics applications, RLZ has also found wider use as a compressor for large text corpora in contexts where random-access support for individual documents is needed~\cite{HPZ11b,TWZ14a,TWZ14b,PMNW15,LPMW16, BCCGSVV2018} and as as a general data compressor~\cite{KVNP20,KL21}. In those contexts, $S_1$ is usually first constructed using substrings sampled from other strings in the collection (Hoobin et al.~\cite{HPZ11b} show that random sampling works well) in a preprocessing phase. The structure of the RLZ parsing reveals a great deal about the repetitive structure of the string collection and several authors have shown that this can be exploited to design efficient compressed indexes for pattern matching~\cite{GGKNP12,DJSS14,NS19}. More recently, the practical utility of RLZ as a more general tool for compressed data structuring has also been demonstrated, compressing suffix arrays~\cite{PZ20,PZ21}, document arrays~\cite{PZ21b} and various components of suffix trees~\cite{FGNPS18}.

\section{Basics}
\label{sec:basics}
Throughout we will consider a {\em string} $S = S[1..n] = S[1]S[2]\ldots S[n]$ on an integer alphabet $\Sigma$ of $\sigma$ symbols. The {\em substring} of $S$ that starts at position $i$ and ends at position $j$, $j \ge i$, denoted $S[i..j]$, is the string $S[i]S[i+1]\ldots S[j]$. If $i > j$, then $S[i..j]$ is the empty string $\varepsilon$. A suffix of $S$ is a substring with ending position $j = n$, and a prefix is a substring with starting position $i = 1$.

\subparagraph{Parsings} A \emph{parsing} of a string $S$ wrt.\ a reference string $R$ is a sequence of substrings of $R$---called phrases---$R[i_1,i_1+l_1-1], R[i_2,i_2+l_2-1],\ldots, R[i_z,i_z+l_z-1]$ such that $S= R[i_1,i_1+l_1-1]\cdot R[i_2,i_2+l_2-1]\cdots R[i_z,i_z+l_z-1]$.
The \emph{encoding} of a parsing consists of the sequence of starting indices and lengths of the phrases $(i_1, l_1), (i_2, l_2), \ldots, (i_z, l_z)$.

The \emph{greedy parsing} of $S$ wrt.\ $R$ is the parsing obtained by processing $S$ from left to right and choosing the longest possible phrase at each step.  For example, let $R =actccta$ and $S =ctctcc$. The greedy parsing of $S$ wrt.\ $R$ gives the phrases $R[2,4] = ctc$, $R[3,5] = tcc$ and the encoding $(2,3),(3,3)$. We can construct the parsing in $O(|R| + |S|)$ time using a suffix tree.


\subparagraph{Relative Lempel-Ziv Compression} 
Throughout the rest of the paper let $\mathcal{S} = \{S_1, S_2,\ldots, S_m\}$ be a collection of $m$ strings of total length $n=\sum_{i=1}^m |S_i|$. The \emph{relative Lempel-Ziv} (RLZ) compression of $\mathcal{S}$ greedily parses each string $S_i$, $i>1$,  wrt.\ $S_1$. The \emph{RLZ compressed representation} of $\mathcal{S}$ then consists of $S_1$ and the encoding of the parsings of each for the strings $S_2, \ldots, S_m$. For each string we also save the number of phrases. In total, compression takes $O(n)$ time. Let $z_R = \sum_{i=2}^m z_i$. The size of the RLZ compression is thus $O(|S_1| + z_R)$. Note that the size depends on the choice of the reference string (i.e.  $S_1$) among the strings in $\mathcal{S}$. To decompress, we decode the phrases of each string using the explicitly stored reference string. This uses $O(\sum_{i=1}^m |S_i|) = O(n)$ time. 

Throughout this paper we use the number of phrases as the measure of compression. In a real compressor, the phrase positions and lengths undergo further processing in order to reduce the total number of bits used by the encoding (see, e.g.,~\cite{FNV13}). We remark that our hierarchical RLZ methods can be trivially adapted to use different encoding costs.

\subparagraph{Graphs} 
Let $G$ be a weighted directed strongly connected graph $G$.  A \emph{spanning arborescence} $A$ of $G$ with root $r$ is a subgraph of $G$ that is a directed rooted tree where all nodes are reachable from $r$.
The weight of an arborescence $S$ is the sum of the weight of the edges in $A$.
A \emph{minimum weight spanning arborescence} (MWSA) $A$ is a spanning arborescence of minimum weight. Note that the root is not fixed in our definition and can thus be any node. For simplicity we have assumed that $G$ is strongly connected in our definition of MWSA since this is always the case in our scenario. Finally, for a node $v$ in a tree, the parent of a node is denoted $\parent(v)$. 

\section{Hierarchical Relative Lempel-Ziv Compression}
\label{sec:hrlz}

Let $\mathcal{S} = \{S_1, S_2, \ldots, S_m \}$ be a collections of strings of total size $n$ as above. 
We construct a rooted tree~$H$, with root $r$, such that each node $v$ represents a unique string $S(v)$ from~$\mathcal{S}$.
The \emph{hierarchical relative Lempel-Ziv} (HRLZ) compression  of $\mathcal{S}$ wrt.~$H$  greedily parses  $S(v)$ wrt.~$S(\parent(v))$ for each non-root node $v$. In total, compression takes $O(n)$ time. The \emph{HRLZ compressed representation} consists of $S(r)$, the edges of $H$, and the encoding of the $m-1$ parsings of the non-root strings. Let $z_H = \sum_{v\in H\setminus \{r\}}z_v$, where $z_v$ is the number of phrases in the parsing of $S(v)$. Thus the size of the HRLZ compression is $O(|H| + |S(r)| + z_H) = O(|S(r)| + z_H)$. Note that the size depends on the choice of tree and assignment of strings from $\mathcal{S}$ to the nodes.  

To decompress, we traverse the tree in breadth first search (BFS) order from the root. We decode the string at each node using the string of the parent node by decoding each phrase. As the string of the parent node is always decoded before or explicitly stored as the root node, this uses $O(\sum_{i=1}^m |S_i|) = O(n)$ time.

\section{Constructing an Optimal Tree}\label{sec:optimaltree}
We first give a simple and inefficient algorithm to construct an optimal tree for the HRLZ compression. The algorithm forms the basis of our efficient algorithm in the following section. Recall that the collection $\mathcal{S} = \{S_1, S_2, \ldots, S_m \}$ consists of $m$ strings of total size~$n$. The algorithm proceeds as follows: 

\medskip

\paragraph{Step 1: Construct Cost Graph}
We first construct a complete weighted directed graph $G$ with $m$ nodes numbered $\{1, \ldots, m\}$ called the \emph{cost graph} of $\mathcal{S}$. Node $i$ corresponds to the string $S_i$ in $\mathcal{S}$ and the weight of edge $(i, j)$ is number of phrases in the greedy parsing of $S_j$ wrt. $S_i$.

We have that $G$ contains $m$ nodes and $m^2$ edges. Computing the weight of edge $(i,j)$ takes $O(|S_j|)$ time. Thus in total we use $O((m-1)\sum_j |S_j|) = O(nm)$ time. The space is $O(m^2)$.

\medskip

\paragraph{Step 2: Construct Minimum Weight Spanning Arborescence}
We then construct a MWSA $A$ of the cost graph $G$ using the algorithm by Tarjan~\cite{Tarjan1977,Camerini1979}. This uses $O(e\log{m})= O(m^2\log{m})$  time. Here $e$ denotes the number of edges in the graph. 

\medskip

\paragraph{Step 3: Construct Compressed Representation}
Finally, we construct the HRLZ compressed from the MWSA $A$. This uses $O(n)$ time. 

\medskip

In total the algorithm uses $O(mn)$ time and $O(m^2)$ space. Note that the algorithm constructs an optimal tree but not necessarily the optimal HRLZ compression since the HRLZ compression also needs to explicitly encode the string of root of the tree. It is straightforward to include the cost of encoding the root string in the algorithm, by adding an additional virtual root $s$ and adding edges $(s, i)$ to every other node $i$, $1 \leq i \leq m$, with weight $|S_i|$. The MWSA of the new graph $G'$ will be rooted in the $s$ and the unique edge out of $s$ determines the optimal root string for HRLZ compression. While $G'$ is not strongly connected, the MWSA is still well-defined and  the MWSA algorithm produces the correct result in the same complexity. In practice, our data sets consists of very similar length strings and hence we have chosen not to  implement this extension.

\section{Sparsifying the Cost Graph via Locality Sensitive Hashing}\label{sec:lsh}
The main bottleneck in the simple algorithm from Section~\ref{sec:optimaltree} is the construction of the complete cost graph in step 1. In this section, we show how to efficiently sparsify the graph using locality sensitive hashing. 

We first construct a sparse subgraph $\overline{G}$ of the complete cost graph. We do this in rounds as follows. Initially, we set $\overline{G}$ to be the graph with $m$ nodes and no edges, and $R = \mathcal{S}$. We repeat the following step until $\overline{G}$ is strongly connected. 
\medskip

\paragraph{Step 1: Generate fingerprints} We first generate fingerprints for each string in $R$ using locality sensitive hashing. Our locality sensitive hashing scheme is based on $k$-mers combined with min-hashing. More precisely, given parameters $k$ and $q$ we pick $q$ hash functions $h_1, \ldots, h_q$ and hash each $k$-mer of each string $S$ in $R$. The fingerprint of $S$ is the sequence $\min_1, \ldots, \min_q$ where $\min_i$ is a minimum value hash obtained with $h_i$. For fast hashing we use the simple \emph{multiply-shift} hashing scheme~\cite{DHKP1997}.

\medskip

\paragraph{Step 2: Generate edges} Let $C$ be a group of strings in $R$ with the same fingerprint. For each ordered pair $(i,j)$ of strings in $C$ we add $(i,j)$ to $\overline{G}$.  

\medskip

\paragraph{Step 3: Pruning $R$} After every $c$'th round for some tuneable parameter $c$ we prune the set $R$ as follows. For every connected component in $\overline{G}$ pick the string $s$ that has had the most number of collisions until now (the total number of collisions of a string $s$ is equal to the sum of the size of the buckets it has been in). We then continue with $R$ being the set of representatives. 

\medskip

Finally, we compute the weight of the edges of the strongly connected graph $\overline{G}$, i.e., for each edge $(i,j)$ we compute the number of phrases in the greedy parsing of $S_j$ wrt. $S_i$. 

\medskip

The computed cost graph is likely to be sparse and thus step 1 and step 2 of the algorithm from  Section~\ref{sec:optimaltree} will be much faster, leading to a much faster solution. Note that the constructed tree is no longer guaranteed to be optimal. We show experimentally in Section~\ref{sec:experiments} that the size of the compression in nearly all cases is within 5\% of optimal.

\section{Speeding Up the Minimum Weight Spanning Arborescense Algorithm}\label{sec:fastarbor}
We now show how to efficiently implement step 2 of the algorithm from Section~\ref{sec:optimaltree} on the sparse cost graph computed in Section~\ref{sec:lsh}. 

Let $\overline{G}$ be the sparse cost graph with $m$ nodes and $e$ edges computed in Section~\ref{sec:lsh}. The MWSA algorithm  by Tarjan~\cite{Tarjan1977,Camerini1979} uses $m$ priority queues $Q_1, Q_2, \ldots, Q_m$, one for each node, were $Q_i$ consists of all edges going into node $v_i$. The queues support the following operations:
\begin{itemize}
\item \textsf{init}$(L)$: Constructs a queue $Q$ containing all the elements in the list $L$.
\item \textsf{extract-min}$(Q_i)$: Returns and removes the minimum element in the queue.
\item \textsf{add}$(Q_i, c)$: Adds a constant $c$ to the value of all elements in the queue.
\item \textsf{meld}$(Q_i, Q_j)$: Adds the elements from $Q_j$ into $Q_i$.
\end{itemize}
The MWSA algorithm uses a \emph{pairing heap}~\cite{FSST86}  to support \textsf{init} in $O(|L|)$ time and the other operations in $O(\log m)$ time. The algorithm uses $O(m)$ \textsf{meld} and \textsf{init} operations, $O(e)$ \textsf{add} and \textsf{extract-min} operations, and the total length of the lists for the \textsf{init} operations is $O(e)$. Thus, the total run time of the queue operations in the MWSA algorithm is $O(e\log{m})$, and this is also  total runtime of the MWSA algorithm.

We present a simple and practical alternative to the pairing heap that we call a \emph{two-level heap}. Our two-level heap leads to a slightly worse theoretical bound of $O(e\log m + m\log^2{m})$ time for the MWSA algorithm. However, we have found that our implementation outperforms the pairing heap in practice. 

The two-level heap consists of a \emph{top heap} $t$ and a list of $q$ \emph{bottom heaps} $B=\{b_1, b_2, \ldots b_q\}$. All heaps are implemented using standard binary heaps~\cite{williams1964}. Each heap $h$ has an associated \emph{offset} $o_h$, such that any stored element $x$ in $h$ represents that actual value $x + o_h$. The top heap $t$ consists of the minimum element in each bottom heap $b\in B$. For each element in the top heap we also store which botton heap it is from. We implement each of the operations as follows: 
\paragraph{\textsf{init}($L$).}
We construct a two-level heap consisting of a single bottom heap $B = \{b\}$ containing the elements of $L$ and a top heap $t$ containing the minimum element of $b$. We set the offsets $o_b$ and $o_t$ of $b$ and $t$, respectively, to be $0$. This uses $O(|L|)$ time and hence the total time for \textsf{init} in the MWSA algorithm is $O(e)$


\paragraph{\textsf{extract-min}($Q_i$).}
We extract the minimum element $x$ from the top heap $t$ and return $x + o_t$. Let $b$ be bottom heap that stored $x$. We extract $x$ from $b$, find the new minimum element $y$ in $b$, and copy $y$ into the top heap with offset $o_b$. This uses $O(\log m)$ time and hence the total time for \textsf{extract-min} in the MWSA algorithm is $O(e\log m)$. 




\paragraph{\textsf{add}$(Q_i, c)$.}
We add $c$ to the offset of the top heap~$t$, i.e., we set $o_t = o_t + c$. This takes constant time and hence the total time for \textsf{add} in the MWSA algorithm is $O(e)$.

\paragraph{\textsf{meld}($Q_i, Q_j$).}
Let $Q_i = (t_i, B_i)$ and $Q_j = (t_j, B_j)$ be the two-level heaps that we want to meld. Let $|B_i|$ and $|B_j|$ be the number of bottom heaps associated with two-level heap $Q_i$ and $Q_j$, respectively, and assume wlog.\ that $|B_i|\geq |B_j|$.
We move each bottom heap $b\in B_j$ into $B_i$, insert the minimum element of $b$ into $t_i$ with offset $o_b + o_{t_j} - o_{t_i}$, and update the offset associated with $b$ to $o_b=o_b + o_{t_j} - o_{t_i}$.


Each time an element in a bottom heap $b$ is moved, we must insert the minimum element of $b$ into a top heap using $O(\log{m})$ time. We only move the bottom heaps of the two-level heap with the fewest bottom heaps and hence the number of times a bottom heap can be moved is $O(\log{m})$. It follows that total time for \textsf{meld} in the MWSA algorithm is $O(m\log^2{m})$. 

\medskip

In total the MSWA algorithm implemented with the two-level heap uses $O(e\log m + m\log^2{m})$ time. 



\section{Experimental Results}\label{sec:experiments}

We implemented the methods for building hierarchical references  described in the previous sections and measured their performance on real biological data. 

\subsection{Implementation details} 
In our experiments, we used $k$-mers of $256$ characters in size and the number of hash functions $q=4$.
We pruned the set $R$ every $c=10$ rounds.
If a group of strings $C$ with identical fingerprints is larger than $2\cdot \sqrt{m}$, we ignore it to avoid adding too many edges to the graph $G$.
Furthermore if $|R|\leq 2\cdot \sqrt{m}$ then for each ordered $(i,j)$ of strings in $R$ we add $(i,j)$ to $G$, and thus finishing the process of construction $G$.

\subsection{Setup}
Experiments were run on Nixos 21.11 kernel version 5.10.115. The compiler was {\tt g++} version 11.3.0 with {\tt -Wall -Wextra -pedantic -O3 -funroll-loops -DNDEBUG -fopenmp -std=gnu++20} options. OpenMP version 4.5 was used to compute the string fingerprints in parallel and compute the edge weights on the cost graph. The CPU was an AMD Ryzen 3900X 12 Core CPU clocked at 4.1 GHz with L1, L2 caches of size 64KiB, 512KiB, per core respectively and a shared L3 cache of size 64MiB. The system had 32GiB of DDR4 3600 MHz memory.  We recorded the CPU wall time using GNU time and {\tt C++} chrono library. Source code is available on request.

\subsection{Data Sets}

\begin{table*}[ht]
\centering
\begin{tabular}{llrrrr}
\hline
\multicolumn{1}{l}{Name} &
\multicolumn{1}{l}{Description}&
\multicolumn{1}{c}{$\sigma$}& 
\multicolumn{1}{c}{$n$}& 
\multicolumn{1}{c}{$m$} &
\multicolumn{1}{c}{$n/m$}\\ \hline
{\em chr19} & Human chromosome 19 assemblies & 5 & $59,125,151,874$ & $1,000$ & $59,125,151$\\
{\em E.coli} & {\em E.coli} genomes & 4 & $1,130,374,882$ & $219$ & $5,161,529$\\ 
{\em SARS-CoV2} & Covid-19 genomes & 5 & $11,949,531,820$ & $400,000$ & $29,873$\\
\hline
\end{tabular}
\caption{Data sets used in experimentation. Columns labelled $\sigma$, $n$, and $m$, give the alphabet size, total collection size, and number of sequences, respectively. The final column shows the average sequence length, for convenience.}\label{table:datasets}
\end{table*}

We evaluated our method using 2,048 copies of human chromosome 19 from the 1000 Genomes Project~\cite{1000genomes}; 10,000 {\em E. coli} genomes taken from the GenomeTrakr project~\cite{genometrakr}, and 400,000 SARS-CoV2 genomes from EBI’s COVID-19 data portal~\cite{covidDataset}. See Table~\ref{table:datasets} for a brief summary of the datasets.

\subsection{Methods Tested}
We included the following methods in our experimental evaluation.

\medskip

\noindent {\bf RLZ.} This corresponds to standard, single reference RLZ. Because the choice of reference can affect overall compression, we report results across a large number of reference selections (further details below).

\medskip

\noindent {\bf Optimal HRLZ.} This is the method described in Section~\ref{sec:optimaltree}, i.e., optimal hierarchical RLZ making use of full weight information.

\medskip

\noindent {\bf Approximate HRLZ.} The LSH variant of hierarchical RLZ described in Section~\ref{sec:lsh}.
    
\medskip

\noindent {\bf LZ.} As a compression baseline, we also compute the full LZ77 parsing of our data sets using the KKP-SE external memory algorithm and software of~\cite{KKP14}\footnote{Code available at \url{https://www.cs.helsinki.fi/group/pads/em_lz77.html}.}. Because it allows phrases to have their source at any previous position in the collection, computing the LZ77 parsing is more computationally demanding than RLZ parsing, and so we compute it only for some prefixes of the collections. For similar reasons, although in principle the above RLZ-based methods could attain parsings as small as the LZ77 parsing, we expect them not to. 

\subsection{Compression Performance}
In this section we compare the number of phrases generated by single reference RLZ to Optimal HRLZ and Approximate HRLZ on the datasets with LZ as a baseline. We also compare the compression time of RLZ and both variants of HRLZ.
    
In preliminary experiments we observed that the number of phrases generated by RLZ depends greatly on the particular reference sequence selected. In the experiments we present here therefore, we ran RLZ multiple times with a number different references and recorded the mean, standard deviation, minimum and maximum number of phrases generated (as well are mean and standard deviation in compression time) for RLZ across those reference sequences. In particular, we applied RLZ the 50, 1000, 100 different reference sequences from {\em E. coli}, {\em SARS-CoV2} and the {\em chr19} dataset, respectively.

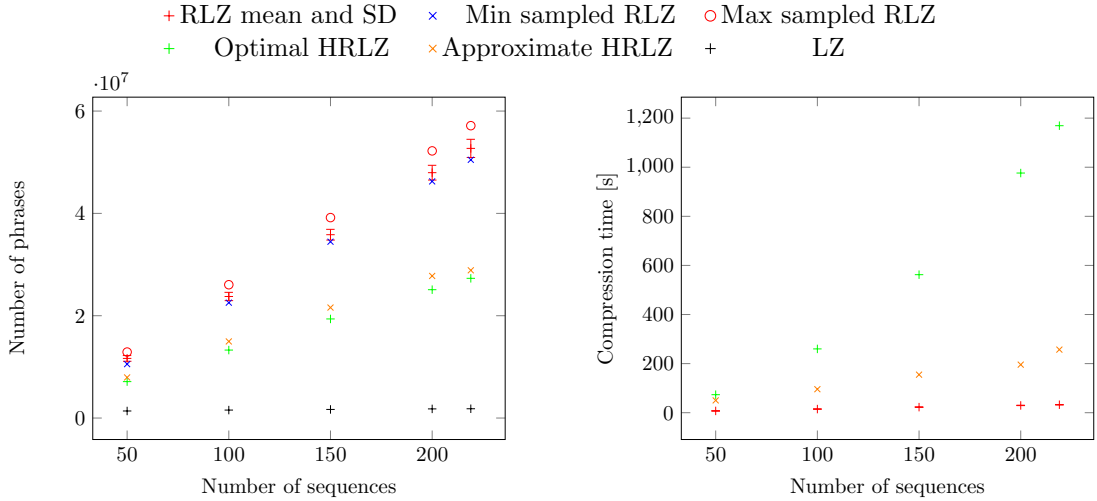
\begin{figure*}[h!]
    \centering
    \begin{tikzpicture}
    \begin{customlegend}[legend entries={RLZ mean and SD\phantom{00},Min sampled RLZ,Max sampled RLZ,Optimal HRLZ, Approximate HRLZ\phantom{00}, LZ}, legend columns=3, legend style={draw=none}]
    \addlegendimage{color=red, only marks, mark=+}
    \addlegendimage{color=blue, only marks, mark=x}
    \addlegendimage{color=red, only marks, mark=o}
    \addlegendimage{color=green, only marks, mark=+}
    \addlegendimage{color=orange, only marks, mark=x}
    \addlegendimage{color=black, only marks, mark=+}
    \end{customlegend}
    \end{tikzpicture}
    
    \begin{tikzpicture}[scale=0.8]
    \begin{axis}[
    xlabel={Number of sequences},
    ylabel={Number of phrases}
    ]
    \addplot[color=red, only marks,  mark=+, error bars/.cd, y dir=both, y explicit]
        table[x index=0, y index=5, y error index=6, col sep=comma] {data/ecoli/phrases.csv};
        
    \addplot[color=blue, only marks,  mark=x]
        table[x index=0, y index=7, col sep=comma] {data/ecoli/phrases.csv};
            
    \addplot[color=red, only marks,  mark=o]
        table[x index=0, y index=8, col sep=comma] {data/ecoli/phrases.csv};
    
    \addplot[color=green, only marks,  mark=+]
        table[x index=0, y index=10, col sep=comma] {data/ecoli/phrases.csv};
    
    \addplot[color=orange, only marks,  mark=x]
        table[x index=0, y index=12, col sep=comma] {data/ecoli/phrases.csv};
    
    \addplot[color=black, only marks,  mark=+]
        table[x index=0, y index=2, col sep=comma] {data/ecoli/phrases.csv};
    
    \end{axis}
    \end{tikzpicture}%
    \hspace{1cm}%
    {
    \begin{tikzpicture}[scale=0.8]
    \begin{axis}[
    xlabel={Number of sequences},
    ylabel={Compression time [s]},
    ]
    \addplot[color=red, only marks,  mark=+, error bars/.cd, y dir=both, y explicit] table[x index=0, y index=3, y error index=4, col sep=comma] {data/ecoli/compress_time.csv};
        
    \addplot[color=green, only marks,  mark=+]
        table[x index=0, y index=5, col sep=comma] {data/ecoli/compress_time.csv};
    
    \addplot[color=orange, only marks,  mark=x]
        table[x index=0, y index=7, col sep=comma] {data/ecoli/compress_time.csv};
    
    \end{axis}
    \end{tikzpicture}
    }
    \caption{Number of phrases generated and the compression time for each algorithm as a function of the number of sequences on the {\em E. coli} dataset.}
    \label{fig:ecoli:phrases}
\end{figure*}

\begin{figure*}[h!]
    \centering    \begin{tikzpicture}
    \begin{customlegend}[legend entries={RLZ mean and SD\phantom{00},Min sampled RLZ,Max sampled RLZ,Optimal HRLZ, Approximate HRLZ\phantom{00}, LZ}, legend columns=3, legend style={draw=none}]
    \addlegendimage{color=red, only marks, mark=+}
    \addlegendimage{color=blue, only marks, mark=x}
    \addlegendimage{color=red, only marks, mark=o}
    \addlegendimage{color=green, only marks, mark=+}
    \addlegendimage{color=orange, only marks, mark=x}
    \addlegendimage{color=black, only marks, mark=+}
    \end{customlegend}
    \end{tikzpicture}
    
    \begin{tikzpicture}[scale=0.70]
    \begin{axis}[
    xlabel={Number of sequences},
    ylabel={Number of phrases},
    legend style={at={(0.5,1)},anchor=south}
    ]
    \addplot[color=red, only marks,  mark=+, error bars/.cd, y dir=both, y explicit]
        table[x index=0, y index=5, y error index=6, col sep=comma] {data/covid/phrases.csv};
        
    \addplot[color=blue, only marks,  mark=x]
        table[x index=0, y index=7, col sep=comma] {data/covid/phrases.csv};
        
    \addplot[color=red, only marks,  mark=o]
        table[x index=0, y index=8, col sep=comma] {data/covid/phrases.csv};
    
    \addplot[color=green, only marks,  mark=+]
        table[x index=0, y index=10, col sep=comma] {data/covid/phrases.csv};
    
    \addplot[color=orange, only marks,  mark=x]
        table[x index=0, y index=12, col sep=comma] {data/covid/phrases.csv};
    
    \addplot[color=black, only marks,  mark=+]
        table[x index=0, y index=2, col sep=comma] {data/covid/phrases.csv};
    
    \end{axis}
    \end{tikzpicture}%
    \begin{tikzpicture}[scale=0.70]
    \begin{axis}[
    xlabel={Number of sequences},
    ]
    \addplot[color=blue, only marks,  mark=x]
        table[x index=0, y index=7, col sep=comma] {data/covid/phrases.csv};
        
    \addplot[color=green, only marks,  mark=+]
        table[x index=0, y index=10, col sep=comma] {data/covid/phrases.csv};
    
    \addplot[color=orange, only marks,  mark=x]
        table[x index=0, y index=12, col sep=comma] {data/covid/phrases.csv};
    
    \addplot[color=black, only marks,  mark=+]
        table[x index=0, y index=2, col sep=comma] {data/covid/phrases.csv};
    
    \end{axis}
    \end{tikzpicture}%
    \begin{tikzpicture}[scale=0.70]
    \begin{axis}[
    xlabel={Number of sequences},
    ylabel={Compression time [s]},
    ]
    \addplot[color=red, only marks,  mark=+, error bars/.cd, y dir=both, y explicit] table[x index=0, y index=2, y error index=3, col sep=comma] {data/covid/compress_time.csv};
        
    \addplot[color=green, only marks,  mark=+]
        table[x index=0, y index=4, col sep=comma] {data/covid/compress_time.csv};
    
    \addplot[color=orange, only marks,  mark=x]
        table[x index=0, y index=5, col sep=comma] {data/covid/compress_time.csv};
    
    \end{axis}
    \end{tikzpicture}
    \caption{Number of phrases generated (left and center) and the compression time (right) for each algorithm as a function of the number of sequences on the SARS-CoV2 dataset. In the center plot we have removed RLZ mean and SD and the Maximum sampled RLZ in order to more clearly display the points for the other series.}
    \label{fig:covid:phrases}
\end{figure*}
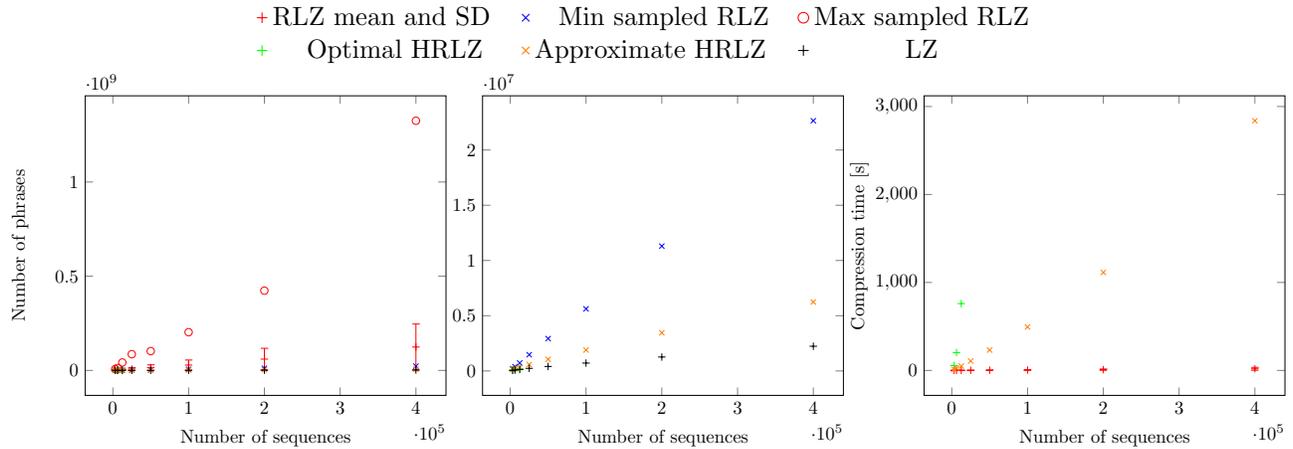

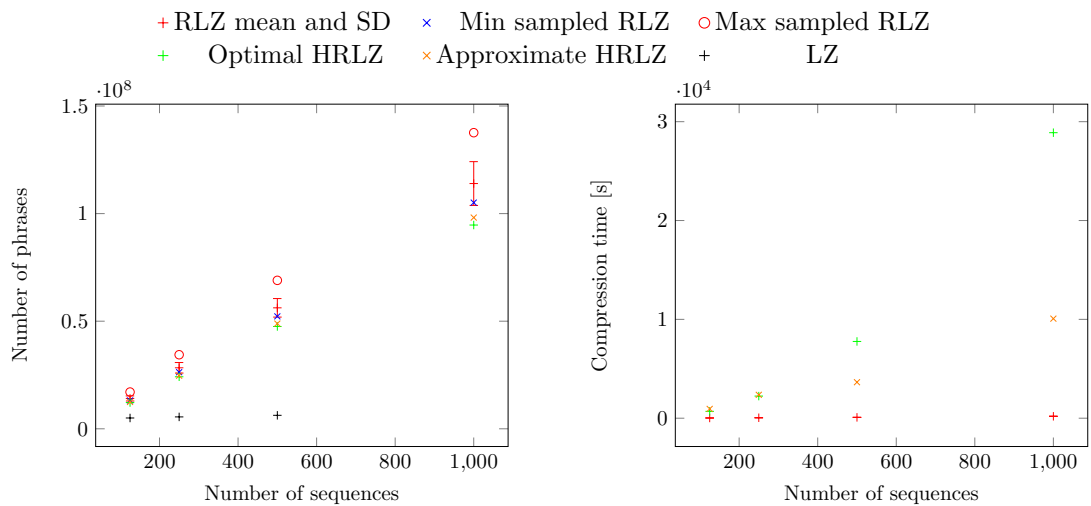
\begin{figure*}[h!]
    \centering
    \begin{tikzpicture}
    \begin{customlegend}[legend entries={RLZ mean and SD\phantom{00},Min sampled RLZ,Max sampled RLZ,Optimal HRLZ, Approximate HRLZ\phantom{00}, LZ}, legend columns=3, legend style={draw=none}]
    \addlegendimage{color=red, only marks, mark=+}
    \addlegendimage{color=blue, only marks, mark=x}
    \addlegendimage{color=red, only marks, mark=o}
    \addlegendimage{color=green, only marks, mark=+}
    \addlegendimage{color=orange, only marks, mark=x}
    \addlegendimage{color=black, only marks, mark=+}
    \end{customlegend}
    \end{tikzpicture}
    
    \begin{tikzpicture}[scale=0.8]
    \begin{axis}[
    xlabel={Number of sequences},
    ylabel={Number of phrases},
    legend style={at={(0.5,1)},anchor=south}
    ]
    \addplot[color=red, only marks,  mark=+, error bars/.cd, y dir=both, y explicit]
        table[x index=0, y index=5, y error index=6, col sep=comma] {data/human_chromosome_19/phrases.csv};
        
    \addplot[color=blue, only marks,  mark=x]
        table[x index=0, y index=7, col sep=comma] {data/human_chromosome_19/phrases.csv};
        
    \addplot[color=red, only marks,  mark=o]
        table[x index=0, y index=8, col sep=comma] {data/human_chromosome_19/phrases.csv};
    
    \addplot[color=green, only marks,  mark=+]
        table[x index=0, y index=10, col sep=comma] {data/human_chromosome_19/phrases.csv};
    
    \addplot[color=orange, only marks,  mark=x]
        table[x index=0, y index=12, col sep=comma] {data/human_chromosome_19/phrases.csv};
    
    \addplot[color=black, only marks,  mark=+]
        table[x index=0, y index=2, col sep=comma] {data/human_chromosome_19/phrases.csv};
    
    \end{axis}
    \end{tikzpicture}%
    \hspace{1cm}%
    \begin{tikzpicture}[scale=0.8]
    \begin{axis}[
    xlabel={Number of sequences},
    ylabel={Compression time [s]},
    ]
    \addplot[color=red, only marks,  mark=+, error bars/.cd, y dir=both, y explicit] table[x index=0, y index=2, y error index=3, col sep=comma] {data/human_chromosome_19/compress_time.csv};
        
    \addplot[color=green, only marks,  mark=+]
        table[x index=0, y index=4, col sep=comma] {data/human_chromosome_19/compress_time.csv};
    
    \addplot[color=orange, only marks,  mark=x]
        table[x index=0, y index=5, col sep=comma] {data/human_chromosome_19/compress_time.csv};
    
    \end{axis}
    \end{tikzpicture}
    \caption{Number of phrases generated and the compression time for each algorithm as a function of the number of sequences on the human chromosome 19 dataset.}
    \label{fig:chr19:phrases}
\end{figure*}

Results are shown in Figures~\ref{fig:ecoli:phrases}-\ref{fig:chr19:phrases}.
We observe that, as expected, as the number of sequences increases the compression rate of both variants of HRLZ improves compared to RLZ, since HRLZ can change the reference used to compress a sequence while RLZ has to use the same reference for every sequence. The biggest difference---roughly a factor of 2---between the best single reference RLZ parsing size and that of the HRLZ variants is on the {\em E.coli} dataset. Differences between the mean single reference RLZ result and the other methods are more modest on {\em SARS-CoV2} and {\em chr19}, however, the plots show a consistently large difference between the worst (maximum) single reference parsing size and the HRLZ variants, emphasising just how badly compression performance can deteriorate for the single reference case if an ill-fitting reference happens to be chosen. We remark that LZ always provides the smallest parsing sizes, but was slower to run than approximate HRLZ, and, for larger datasets, required more disk space ($>$1TB) than our system had available. 

The number of phrases generated by the approximate HRLZ variant is very close to that of the optimal HRLZ algorithm algorithm---within 5\% on all runs. Moreover, we see that, as intended, compression time for the approximate HRLZ variant is significantly faster than that of optimal HRLZ on all the datasets---around 6 times faster on {\em E.coli}, three times faster on chromosome 19, and orders of magnitude faster on {\em SARS-CoV2}, due to the large number of sequences in that set.

While this improvement in compression time of approximate HRLZ over optimal HRLZ is pleasing, Figure~\ref{fig:covid:phrases} clearly shows that there is a significant overhead to deriving a good hierarchy---even an approximate one---and that the compression time of approximate HRLZ grows faster than single reference RLZ. We remark, however, that compression is often a once-off process, and that in most contexts (including genomic applications), having rapid decompression is much more important, an aspect we examine next.

\subsection{Decompression Time}

\begin{figure*}[h!]
    \centering
    \begin{tikzpicture}
    \begin{customlegend}[legend entries={RLZ\phantom{00}, Optimal HRLZ\phantom{00}, Approximate HRLZ}, legend columns=3, legend style={draw=none}]
    \addlegendimage{color=blue, only marks, mark=x}
    \addlegendimage{color=orange, only marks, mark=x}
    \addlegendimage{color=black, only marks, mark=+}
    \end{customlegend}
    \end{tikzpicture}
    
    \begin{tikzpicture}[scale=0.7]
    \begin{axis}[
    xlabel={Number of sequences},
    ylabel={Decompression time [s]}
    ]
    \addplot[color=blue, only marks,  mark=x]
        table[x index=0, y index=1, col sep=comma] {data/ecoli/decompress_time.csv};
    
    \addplot[color=orange, only marks,  mark=x]
        table[x index=0, y index=2, col sep=comma] {data/ecoli/decompress_time.csv};
    
    \addplot[color=black, only marks,  mark=+]
        table[x index=0, y index=3, col sep=comma] {data/ecoli/decompress_time.csv};
    
    \end{axis}
    \end{tikzpicture}%
    \begin{tikzpicture}[scale=0.7]
    \begin{axis}[
    xlabel={Number of sequences},
    legend style={at={(0.5,1)},anchor=south}
    ]
    \addplot[color=blue, only marks,  mark=x]
        table[x index=0, y index=1, col sep=comma] {data/covid/decompress_time.csv};
    
    \addplot[color=orange, only marks,  mark=x]
        table[x index=0, y index=2, col sep=comma] {data/covid/decompress_time.csv};
    
    \addplot[color=black, only marks,  mark=+]
        table[x index=0, y index=3, col sep=comma] {data/covid/decompress_time.csv};
    
    \end{axis}
    \end{tikzpicture}%
    \begin{tikzpicture}[scale=0.7]
    \begin{axis}[
    xlabel={Number of sequences}
    ]
    \addplot[color=blue, only marks,  mark=x]
        table[x index=0, y index=1, col sep=comma] {data/human_chromosome_19/decompress_time.csv};
    
    \addplot[color=orange, only marks,  mark=x]
        table[x index=0, y index=2, col sep=comma] {data/human_chromosome_19/decompress_time.csv};
    
    \addplot[color=black, only marks,  mark=+]
        table[x index=0, y index=3, col sep=comma] {data/human_chromosome_19/decompress_time.csv};
    
    \end{axis}
    \end{tikzpicture}%
    \caption{Decompression time as a function of the number of sequences in the {\tt E. coli} (left), SARS-CoV2 (center) and human chromosome 19 dataset (right) dataset.}
    \label{fig:decompress}
\end{figure*}
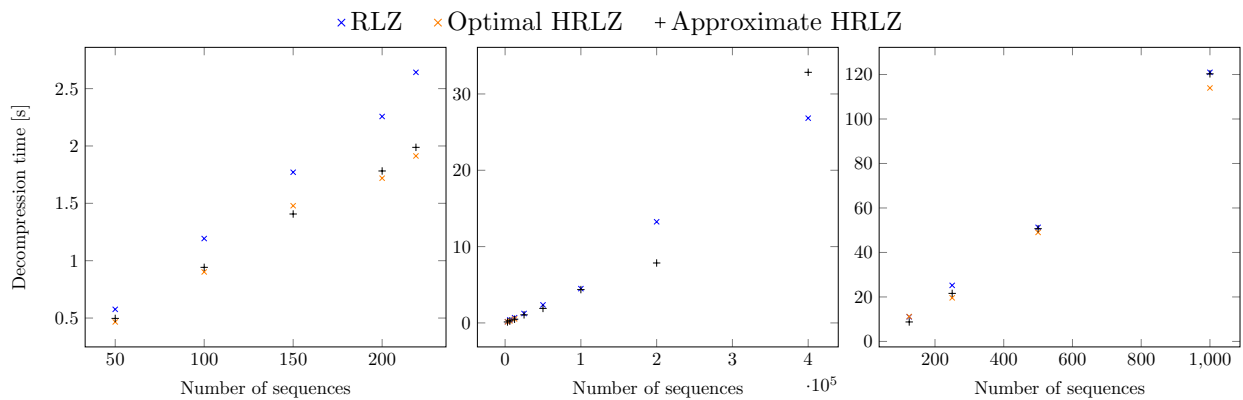

Figure~\ref{fig:decompress} displays decompression times for the various methods. The experiment measured the time to decompress the full data set and write it to disk. This kind of streaming decompression use case is typical for, e.g., multi-pass index construction, machine learning, and data mining processes~\cite{FM10}. All methods have to write the same amount of data to storage when decompressing. Our HRLZ variants decompress sequences in BFS order according to hierarchy imposed on the sequences. This may require sequences that have previously written to disk being read back into memory (at most once) when they are needed as a reference in the decompression of other sequences.

RLZ and the HRLZ variants have similar decompression performance characteristics.
Interestingly, on the {\tt E. coli} dataset both variants of HRLZ outperformed RLZ in decompression time. We believe this is because the longer phrases produced by the HRLZ variants result in fewer cache misses. On larger data sets, which approach the size of RAM on the test machine, RLZ and HRLZ have similar decompression times, which RLZ being occasionally marginally faster (on the largest of the {\em E.coli} data sets, for example). This can be explained by the above mentioned need for the HRLZ variants to read previously written sequences back into memory (when they are needed as reference sequences). Whenever the whole decompressed data set fits comfortably in RAM (32GB on our test machine), operating system caching is likely to make reads almost free, as no disk seek will be required. However, on larger data sets, reads more frequently result in disk seeks, slowing decompression. We observed that maximium node depth on all data sets was relatively low: 31, 36, and 206 for {\em chr19}, {\em E.coli}, and {\em SARS-CoV2}, respectively (for Approximate HRLZ). This suggests it may be profitable to use a DFS decompression strategy, as the working set will remain relatively small---something we will explore in future work. 

\section{Concluding Remarks}
\label{sec:conclusion}

We have shown that traditional, single-reference RLZ compression can be significantly outperformed by imposing a hierarchy on the sequences to be compressed using a sequence's parent in the hierarchy as its reference sequence. Moreover, we have described efficient methods by which hierarchies can be efficiently obtained. Our experiments show that the time to subsequently decompress the set of sequences are at worst negligibly slower, and sometimes even faster than the single reference baseline. 

There are many directions future work could take. 
Apart from compression, another feature of RLZ that makes it attractive in a genomic context is its ready support for efficient random access to individual sequences (and indeed substrings): having a compact, easily accessible representation of the genome sequences compliments popular indexing methods that do not readily support random access themselves (e.g.,~\cite{MKBGLM20}). Supporting random access (and indeed indexed search) to HRLZ compressed data is an interesting avenue for future work.

 As noted in the Introduction, several recent works have demonstrated the practical utility of using RLZ as a tool for compressed data structuring~\cite{FGNPS18,PZ20,PZ21,PZ21b}. In that context, an artificial reference sequence is usually constructed from repeated pieces (e.g., subarrays, or subtrees) of the data structure to be compressed. It would be interesting to see if our methods could be adapted to construct better artificial reference sequences for use in those scenarios.
 
 Finally, in this work we have used purely algorithmic methods to derive hierarchies for datasets: no biological characteristics of the sequences have been used. However, the field of phylogenetics has developed many techniques for imposing a hierarchy on a set of individuals based on biologically meaningful features in their genomic content. It may be interesting to examine any similarities between phylogenetic trees and our RLZ-based hierarchies, and whether phylogenetic trees may serve as good hierarchies in the context of compression.

\newpage

\bibliographystyle{abbrv}
\bibliography{rlz}

\begin{thebibliography}{10}

\bibitem{covidDataset}
Coronavirus genomes – {NCBI} datasets.
\newblock Accessed 18/05/2022,
  \url{https://www.ncbi.nlm.nih.gov/datasets/coronavirus/genomes/}.

\bibitem{BCCGSVV2018}
P.~Bille, A.~R. Christiansen, P.~H. Cording, I.~L. G{\o}rtz, F.~R.
  Skjoldjensen, H.~W. Vildh{\o}j, and S.~Vind.
\newblock Dynamic relative compression, dynamic partial sums, and substring
  concatenation.
\newblock {\em Algorithmica}, 80(11):3207--3224, 2018.

\bibitem{BG2020}
P.~Bille and I.~L. G{\o}rtz.
\newblock Random access in persistent strings.
\newblock In {\em Proc. 31st ISAAC}, 2020.

\bibitem{Camerini1979}
P.~M. Camerini, L.~Fratta, and F.~Maffioli.
\newblock A note on finding optimum branchings.
\newblock {\em Networks}, 9(4):309--312, 1979.

\bibitem{DDG13}
S.~Deorowicz, A.~Danek, and S.~Grabowski.
\newblock Genome compression: a novel approach for large collections.
\newblock {\em Bioinformatics}, 29(20):2572--2578, 2013.

\bibitem{DG11a}
S.~Deorowicz and S.~Grabowski.
\newblock Robust relative compression of genomes with random access.
\newblock {\em Bioinformatics}, 27(21):2979--2986, 2011.

\bibitem{DHKP1997}
M.~Dietzfelbinger, T.~Hagerup, J.~Katajainen, and M.~Penttonen.
\newblock A reliable randomized algorithm for the closest-pair problem.
\newblock {\em J. Algorithms}, 25(1):19--51, 1997.

\bibitem{DJSS14}
H.~H. Do, J.~Jansson, K.~Sadakane, and W.~Sung.
\newblock Fast relative lempel-ziv self-index for similar sequences.
\newblock {\em Theor. Comput. Sci.}, 532:14--30, 2014.

\bibitem{FGNPS18}
A.~Farruggia, T.~Gagie, G.~Navarro, S.~J. Puglisi, and J.~Sir{\'{e}}n.
\newblock Relative suffix trees.
\newblock {\em Comput. J.}, 61(5):773--788, 2018.

\bibitem{FM10}
P.~Ferragina and G.~Manzini.
\newblock On compressing the textual web.
\newblock In {\em Proc. 3rd WSDM}, pages 391--400, 2010.

\bibitem{FNV13}
P.~Ferragina, I.~Nitto, and R.~Venturini.
\newblock On the bit-complexity of {L}empel-{Z}iv compression.
\newblock {\em {SIAM} J. Comput.}, 42(4):1521--1541, 2013.

\bibitem{FSST86}
M.~L. Fredman, R.~Sedgewick, D.~D. Sleator, and R.~E. Tarjan.
\newblock The pairing heap: A new form of self-adjusting heap.
\newblock {\em Algorithmica}, 1(1):111--129, 1986.

\bibitem{GGKNP12}
T.~Gagie, P.~Gawrychowski, J.~K{\"{a}}rkk{\"{a}}inen, Y.~Nekrich, and S.~J.
  Puglisi.
\newblock A faster grammar-based self-index.
\newblock In {\em Proc. 6th LATA}, pages 240--251, 2012.

\bibitem{HPZ11b}
C.~Hoobin, S.~J. Puglisi, and J.~Zobel.
\newblock Relative {Lempel-Ziv} factorization for efficient storage and
  retrieval of web collections.
\newblock {\em Proc. VLDB Endowment}, 5(3):265--273, 2011.

\bibitem{KKP14}
J.~K{\"{a}}rkk{\"{a}}inen, D.~Kempa, and S.~J. Puglisi.
\newblock {L}empel-{Z}iv parsing in external memory.
\newblock In {\em Proc. 24th DCC}, pages 153--162, 2014.

\bibitem{KL21}
D.~Kempa and B.~Langmead.
\newblock Fast and space-efficient construction of {AVL} grammars from the
  {LZ77} parsing.
\newblock In {\em Proc. 29th ESA}, pages 56:1--56:14, 2021.

\bibitem{KVNP20}
D.~Kosolobov, D.~Valenzuela, G.~Navarro, and S.~J. Puglisi.
\newblock Lempel-ziv-like parsing in small space.
\newblock {\em Algorithmica}, 82(11):3195--3215, 2020.

\bibitem{KPZ10}
S.~Kuruppu, S.~J. Puglisi, and J.~Zobel.
\newblock Relative {Lempel-Ziv} compression of genomes for large-scale storage
  and retrieval.
\newblock In {\em Proc. 17th SPIRE}, pages 201--206, 2010.

\bibitem{LPMW16}
K.~Liao, M.~Petri, A.~Moffat, and A.~Wirth.
\newblock Effective construction of relative lempel-ziv dictionaries.
\newblock In {\em Proc. 25th WWW}, pages 807--816, 2016.

\bibitem{MKASHMCHH21}
T.~M{\"a}klin, T.~Kallonen, J.~Alanko, {\O}.~Samuelsen, K.~Hegstad,
  V.~M{\"a}kinen, J.~Corander, E.~Heinz, and A.~Honkela.
\newblock Bacterial genomic epidemiology with mixed samples.
\newblock {\em Microbial Genomics}, 7(11), 2021.

\bibitem{MKBGLM20}
T.~Mun, A.~Kuhnle, C.~Boucher, T.~Gagie, B.~Langmead, and G.~Manzini.
\newblock Matching reads to many genomes with the r-index.
\newblock {\em J. Comput. Biol.}, 27(4):514--518, 2020.

\bibitem{NS19}
G.~Navarro and V.~Sepulveda.
\newblock Practical indexing of repetitive collections using relative
  {L}empel-{Z}iv.
\newblock In {\em Proc. 29th DCC}, pages 201--210, 2019.

\bibitem{NSMG19}
G.~Navarro, V.~Sepulveda, M.~Mar{\'{\i}}n, and S.~Gonz{\'{a}}lez.
\newblock Compressed filesystem for managing large genome collections.
\newblock {\em Bioinformatics}, 35(20):4120--4128, 2019.

\bibitem{PMNW15}
M.~Petri, A.~Moffat, P.~C. Nagesh, and A.~Wirth.
\newblock Access time tradeoffs in archive compression.
\newblock In {\em Proc. 11th AIRS}, pages 15--28, 2015.

\bibitem{PZ20}
S.~J. Puglisi and B.~Zhukova.
\newblock Relative {Lempel}-{Ziv} compression of suffix arrays.
\newblock In {\em Proc. SPIRE}, LNCS 12303, pages 89--96. Springer, 2020.

\bibitem{PZ21b}
S.~J. Puglisi and B.~Zhukova.
\newblock Document retrieval hacks.
\newblock In {\em Proc. 19th SEA}, pages 12:1--12:12, 2021.

\bibitem{PZ21}
S.~J. Puglisi and B.~Zhukova.
\newblock Smaller {RLZ}-compressed suffix arrays.
\newblock In {\em Proc. 31st DCC}, 2021.

\bibitem{genometrakr}
E.~Stevens et~al.
\newblock The public health impact of a publically available, environmental
  database of microbial genomes.
\newblock {\em Front. Microbiol.}, 8(808), 2017.

\bibitem{SS82}
J.~A. Storer and T.~G. Szymanski.
\newblock Data compression via textual substitution.
\newblock {\em J. {ACM}}, 29(4):928--951, 1982.

\bibitem{Tarjan1977}
R.~E. Tarjan.
\newblock Finding optimum branchings.
\newblock {\em Networks}, 7(1):25--35, 1977.

\bibitem{1000genomes}
{The 1000 Genomes Project Consortium}.
\newblock A global reference for human genetic variation.
\newblock {\em Nature}, 526:68--74, 2015.

\bibitem{TWZ14a}
J.~Tong, A.~Wirth, and J.~Zobel.
\newblock Compact auxiliary dictionaries for incremental compression of large
  repositories.
\newblock In {\em Proc. 23rd CIKM}, pages 1629--1638, 2014.

\bibitem{TWZ14b}
J.~Tong, A.~Wirth, and J.~Zobel.
\newblock Principled dictionary pruning for low-memory corpus compression.
\newblock In {\em Proc. 37th SIGIR}, pages 283--292, 2014.

\bibitem{VNVPM18}
D.~Valenzuela, T.~Norri, N.~V{\"{a}}lim{\"{a}}ki, E.~Pitk{\"{a}}nen, and
  V.~M{\"{a}}kinen.
\newblock Towards pan-genome read alignment to improve variation calling.
\newblock {\em {BMC} Genom.}, 19({S2}), 2018.

\bibitem{williams1964}
J.~W.~J. Williams.
\newblock Algorithm 232: heapsort.
\newblock {\em Commun. ACM}, 7:347--348, 1964.

\end{thebibliography}

\clearpage
\appendix

\onecolumn

\section{Additional Figures}

This appendix contains additional plots as well as data used to generate plots in the main document.

\begin{table*}[ht]
    \centering
    \begin{tabular}{|l|l|l|l|l|l|l|l|}
    \hline
        Size & lz   & rlz   mean & rlz phrases SD & rlz   min & rlz   max & opt-hrlz   & approx-hrlz   \\ \hline
        50 & 1379061 & 11655549.5 & 559931.7 & 10538134 & 12897990 & 7119429 & 7937736 \\ \hline
        100 & 1556541 & 23757023.7 & 810846.0 & 22548503 & 26054391 & 13284440 & 14972229 \\ \hline
        150 & 1681574 & 35834782.4 & 1047249.5 & 34465391 & 39177365 & 19369695 & 21595326 \\ \hline
        200 & 1787317 & 47959074.9 & 1440738.5 & 46250237 & 52215699 & 25087136 & 27773606 \\ \hline
        219 & 1815168 & 52703827.2 & 1775967.6 & 50474351 & 57142680 & 27307037 & 28874750 \\ \hline
    \end{tabular}
    \caption{Number of phrases generated for each algorithm as a function of the number of sequences on the {\em E. coli} data set.}
\end{table*}

\begin{table*}[ht]
    \centering
    \begin{tabular}{|l|l|l|l|l|l|l|l|l|}
    \hline
        Size & lz mean & lz SD & rlz mean & rlz SD & opt-hrlz mean & opt-hrlz SD & approx-hrlz mean & approx-hrlz SD \\ \hline
        50 & 29.9 & 0.0 & 7.3 & 0.3 & 73.4 & 0.0 & 50.1 & 0.0 \\ \hline
        100 & 62.2 & 0.0 & 14.8 & 0.4 & 259.9 & 0.0 & 95.7 & 0.0 \\ \hline
        150 & 97.7 & 0.0 & 22.7 & 0.8 & 562.3 & 0.0 & 155.0 & 0.0 \\ \hline
        200 & 134.5 & 0.0 & 29.6 & 0.7 & 976.3 & 0.0 & 195.8 & 0.0 \\ \hline
        219 & 146.8 & 0.0 & 32.5 & 0.8 & 1169.2 & 0.0 & 257.0 & 0.0 \\ \hline
    \end{tabular}
    \caption{Compression time for each algorithm as a function of the number of sequences on the {\em E. coli} data set.}
\end{table*}

\begin{table*}[ht]
    \centering
    \begin{tabular}{|l|l|l|l|l|l|l|l|}
    \hline
        Size & lz  & rlz  mean & rlz  SD & rlz  min & rlz  max & opt-hrlz  & approx-hrlz  \\ \hline
        3125 & 50164 & 1750673.1 & 1217565.0 & 205590 & 7428924 & 82549 & 126276 \\ \hline
        6250 & 81991 & 2466793.8 & 1813057.6 & 390365 & 12676001 & 141909 & 191162 \\ \hline
        12500 & 133149 & 3520189.7 & 3267927.5 & 721806 & 42117838 & 227037 & 310147 \\ \hline
        25000 & 236196 & 7420125.1 & 7371156.9 & 1472814 & 86128962 & nan & 584710 \\ \hline
        50000 & 408189 & 14991942.6 & 14499649.9 & 2928707 & 102551328 & nan & 1055514 \\ \hline
        100000 & 714174 & 28248696.0 & 26879832.7 & 5623338 & 202889182 & nan & 1899434 \\ \hline
        200000 & 1262731 & 59936818.4 & 57547861.2 & 11299760 & 423360338 & nan & 3457888 \\ \hline
        400000 & 2235801 & 124068851.9 & 122098365.7 & 22646367 & 1325084761 & nan & 6240673 \\ \hline
    \end{tabular}
    \caption{Number of phrases generated for each algorithm as a function of the number of sequences on the SARS-CoV2 data set}
\end{table*}

\begin{table*}[ht]
    \centering
    \begin{tabular}{|l|l|l|l|l|l|}
    \hline
        Size & lz & rlz mean & rlz SD & opt-hrlz & approx-hrlz \\ \hline
        3125 & 9.4 & 0.2 & 0.1 & 55.1 & 11.2 \\ \hline
        6250 & 19.2 & 0.3 & 0.2 & 202.3 & 23.4 \\ \hline
        12500 & 43.6 & 0.5 & 0.4 & 759.6 & 49.8 \\ \hline
        25000 & 90.7 & 1.1 & 0.9 & nan & 106.7 \\ \hline
        50000 & 203.1 & 2.2 & 1.8 & nan & 231.8 \\ \hline
        100000 & 460.5 & 4.3 & 3.2 & nan & 494.0 \\ \hline
        200000 & 980.3 & 8.9 & 7.2 & nan & 1114.2 \\ \hline
        400000 & 2200.1 & 18.2 & 16.3 & nan & 2837.9 \\ \hline
    \end{tabular}
    \caption{Compression time for each algorithm as a function of the number of sequences on the SARS-CoV2 data set.}
\end{table*}
 

\begin{table*}[ht]
    \centering
    \begin{tabular}{|l|l|l|l|l|l|l|l|}
    \hline
        Size & lz  & rlz  mean & rlz  SD & rlz  min & rlz  max & opt-hrlz  & approx-hrlz  \\ \hline
        125 & 5024871 & 14099089.1 & 1182233.9 & 13153108 & 17085211 & 12196621 & 12470163 \\ \hline
        250 & 5526491 & 28344818.1 & 2449630.5 & 26299422 & 34412724 & 24174861 & 24688193 \\ \hline
        500 & 6263992 & 56212384.1 & 4308102.0 & 52240721 & 68983696 & 47547689 & 48888371 \\ \hline
        1000 & nan & 113964055.0 & 10176846.9 & 105106697 & 137596020 & 94684015 & 98157745 \\ \hline
    \end{tabular}
    \caption{Number of phrases generated for each algorithm as a function of the number of sequences on the human chromosome 19 data set.}
\end{table*}

\begin{table*}[ht]
    \centering
    \begin{tabular}{|l|l|l|l|l|l|}
    \hline
        Size & lz & rlz mean & rlz SD & opt-hrlz & approx-hrlz \\ \hline
        125 & 1135.5 & 24.1 & 1.4 & 694.7 & 941.9 \\ \hline
        250 & 2528.4 & 45.5 & 2.9 & 2238.0 & 2366.1 \\ \hline
        500 & 6231.0 & 85.6 & 5.2 & 7765.9 & 3641.0 \\ \hline
        1000 & nan & 193.6 & 12.5 & 28886.1 & 10073.1 \\ \hline
    \end{tabular}
    \caption{Compression time for each algorithm as a function of the number of sequences on the human chromosome 19 data set.}
\end{table*}


\begin{table*}[ht]
    \centering
    \begin{tabular}{|l|l|l|l|}
    \hline
        Size & rlz & opt-hrlz & approx-hrlz \\ \hline
        3125 & 0.159 & 0.138 & 0.126 \\ \hline
        6250 & 0.397 & 0.260 & 0.279 \\ \hline
        12500 & 0.680 & 0.522 & 0.483 \\ \hline
        25000 & 1.235 & nan & 1.061 \\ \hline
        50000 & 2.360 & nan & 1.900 \\ \hline
        100000 & 4.492 & nan & 4.366 \\ \hline
        200000 & 13.252 & nan & 7.852 \\ \hline
        400000 & 26.827 & nan & 32.835 \\ \hline
    \end{tabular}
    \caption{Decompression time as a function of the number of sequences in the {\tt E. coli} data set.}
\end{table*}

\begin{table*}[ht]
    \centering
    \begin{tabular}{|l|l|l|l|}
    \hline
        Size & rlz & opt-hrlz & approx-hrlz \\ \hline
        50 & 0.576 & 0.466 & 0.497 \\ \hline
        100 & 1.193 & 0.902 & 0.942 \\ \hline
        150 & 1.771 & 1.479 & 1.407 \\ \hline
        200 & 2.257 & 1.719 & 1.782 \\ \hline
        219 & 2.642 & 1.913 & 1.988 \\ \hline
    \end{tabular}
    \caption{Decompression time as a function of the number of sequences in the SARS-CoV2 (center) data set.}
\end{table*}

\begin{table*}[ht]
    \centering
    \begin{tabular}{|l|l|l|l|}
    \hline
        Size & rlz & opt-hrlz & approx-hrlz \\ \hline
        125 & 11.071 & 11.186 & 8.759 \\ \hline
        250 & 25.200 & 19.592 & 21.600 \\ \hline
        500 & 51.297 & 49.011 & 50.697 \\ \hline
        1000 & 120.984 & 113.942 & 120.307 \\ \hline
    \end{tabular}
    \caption{Decompression time as a function of the number of sequences in the human chromosome 19 data set (right) data set.}
\end{table*}


\begin{table*}[ht]
    \centering
    \begin{tabular}{|l|l|l|l|l|}
    \hline
        Size & opt-hrlz average& opt-hrlz max  & approx-hrlz average  & approx-hrlz max  \\ \hline
        50 & 7.6 & 14 & 5.3 & 10 \\ \hline
        100 & 10.1 & 19 & 8.5 & 17 \\ \hline
        150 & 17.2 & 35 & 10.7 & 21 \\ \hline
        200 & 14.4 & 26 & 10.9 & 19 \\ \hline
        219 & 16.9 & 34 & 16.7 & 37 \\ \hline
    \end{tabular}
    \caption{The average node depth and maximum node depth for the generated rooted tree for HRLZ as a function of the number of sequences in the {\tt E. coli}  data set.}
\end{table*}

\begin{table*}[ht]
    \centering
    \begin{tabular}{|l|l|l|l|l|}
    \hline
        Size & opt-hrlz average  & opt-hrlz max  & approx-hrlz average  & approx-hrlz max \\ \hline
        3125 & 13.6 & 27 & 45.1 & 75 \\ \hline
        6250 & 14.1 & 36 & 34.4 & 60 \\ \hline
        12500 & 16.4 & 36 & 61.0 & 125 \\ \hline
        25000 & nan & nan & 52.6 & 99 \\ \hline
        50000 & nan & nan & 46.5 & 92 \\ \hline
        100000 & nan & nan & 95.8 & 181 \\ \hline
        200000 & nan & nan & 103.8 & 176 \\ \hline
        400000 & nan & nan & 100.2 & 206 \\ \hline
    \end{tabular}
    \caption{The average node depth and maximum node depth for the generated rooted tree for HRLZ as a function of the number of sequences in the  SARS-CoV2 data set.}
\end{table*}

\begin{table*}[ht]
    \centering
    \begin{tabular}{|l|l|l|l|l|}
    \hline
        Size & opt-hrlz average  & opt-hrlz max  & approx-hrlz average  & approx-hrlz max  \\ \hline
        125 & 9.1 & 15 & 17.4 & 26 \\ \hline
        250 & 12.3 & 20 & 12.6 & 26 \\ \hline
        500 & 13.8 & 25 & 20.2 & 35 \\ \hline
        1000 & 22.1 & 42 & 17.9 & 31 \\ \hline
    \end{tabular}
    \caption{The average node depth and maximum node depth for the generated rooted tree for HRLZ as a function of the number of sequences in the  human chromosome 19 data set.}
\end{table*}

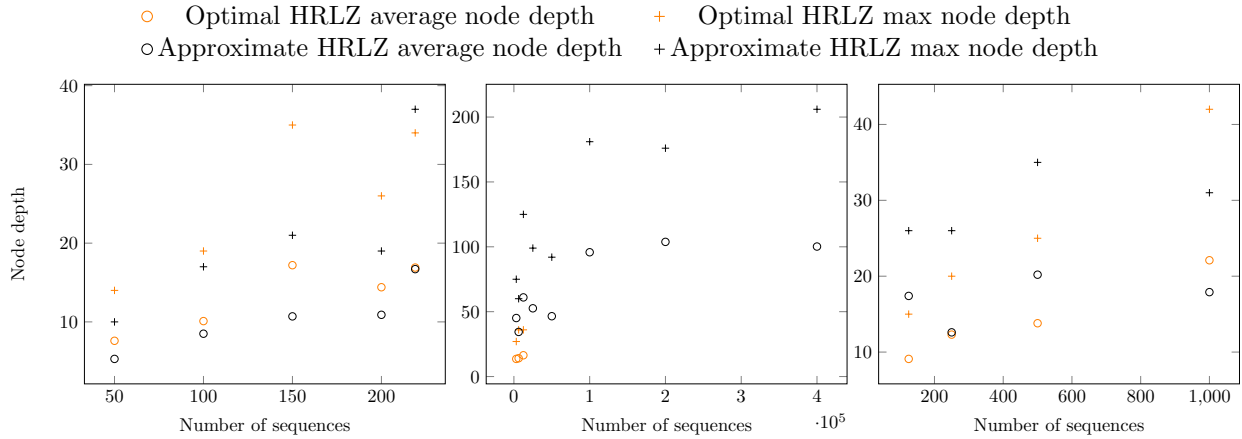
\begin{figure*}[h!]
    \centering
    \begin{tikzpicture}
    \begin{customlegend}[legend entries={Optimal HRLZ average node depth\phantom{00}, Optimal HRLZ max node depth, Approximate HRLZ average node depth\phantom{00}, Approximate HRLZ max node depth}, legend columns=2, legend style={draw=none}]
    \addlegendimage{color=orange, only marks, mark=o}
    \addlegendimage{color=orange, only marks, mark=+}
    \addlegendimage{color=black, only marks, mark=o}
    \addlegendimage{color=black, only marks, mark=+}
    \end{customlegend}
    \end{tikzpicture}
    \begin{tikzpicture}[scale=0.7]
    \begin{axis}[
    xlabel={Number of sequences},
    ylabel={Node depth}
    ]
    
    \addplot[color=orange, only marks,  mark=o]
        table[x index=0, y index=1, col sep=comma] {data/ecoli/node_depth.csv};
        
    \addplot[color=orange, only marks,  mark=+]
        table[x index=0, y index=2, col sep=comma] {data/ecoli/node_depth.csv};
        
    \addplot[color=black, only marks,  mark=o]
        table[x index=0, y index=3, col sep=comma] {data/ecoli/node_depth.csv};
    
    \addplot[color=black, only marks,  mark=+]
        table[x index=0, y index=4, col sep=comma] {data/ecoli/node_depth.csv};
 
    \end{axis}
    \end{tikzpicture}%
    \begin{tikzpicture}[scale=0.7]
    \begin{axis}[
    xlabel={Number of sequences},
    legend style={at={(0.5,1)},anchor=south}
    ]
    \addplot[color=orange, only marks,  mark=o]
        table[x index=0, y index=1, col sep=comma] {data/covid/node_depth.csv};
        
    \addplot[color=orange, only marks,  mark=+]
        table[x index=0, y index=2, col sep=comma] {data/covid/node_depth.csv};
        
    \addplot[color=black, only marks,  mark=o]
        table[x index=0, y index=3, col sep=comma] {data/covid/node_depth.csv};
    
    \addplot[color=black, only marks,  mark=+]
        table[x index=0, y index=4, col sep=comma] {data/covid/node_depth.csv};
    
    \end{axis}
    \end{tikzpicture}%
    \begin{tikzpicture}[scale=0.7]
    \begin{axis}[
    xlabel={Number of sequences}
    ]
    \addplot[color=orange, only marks,  mark=o]
        table[x index=0, y index=1, col sep=comma] {data/human_chromosome_19/node_depth.csv};
        
    \addplot[color=orange, only marks,  mark=+]
        table[x index=0, y index=2, col sep=comma] {data/human_chromosome_19/node_depth.csv};
        
    \addplot[color=black, only marks,  mark=o]
        table[x index=0, y index=3, col sep=comma] {data/human_chromosome_19/node_depth.csv};
    
    \addplot[color=black, only marks,  mark=+]
        table[x index=0, y index=4, col sep=comma] {data/human_chromosome_19/node_depth.csv};
 
    \end{axis}
    \end{tikzpicture}%
    \caption{The average node depth and maximum node depth for the generated rooted tree for HRLZ as a function of the number of sequences in the {\tt E. coli} (left), SARS-CoV2 (center) and human chromosome 19 dataset (right) dataset.}
    \label{fig:nodedepth}
\end{figure*}

\end{document}